**EyeAgent: An Agentic AI System for Multimodal Clinical Decision Support in Ophthalmology**

**Author:**

Danli Shi, MD[1,2#], Xiaolan Chen, MD[1#], Bingjie Yan, MSc[1#], Weiyi Zhang, MSc[1], Jiancheng Yang, PhD[3,4], Ruoyu Chen, MD[1], Xinyuan Wu, MD[1], Siyu Huang, DEng[5], Bowen Liu, MSc[1], Ziyu Gao, MEng[1], Yue Wu, MD[1], Pusheng Xu, MD[1], Senlin Lin, MSc[6], Kai Jin, MD[7], Xia Gong, MD[1], Yih Chung Tham, PhD[8], Xiujuan Zhang, PhD[9], Li Dong, MD[10], Yuzhou Zhang, PhD[11], Jason Yam, PhD[11], Guangming Jin, PhD[12], Xiaohu Ding, MD[12], Haidong Zou, MD[6], Yalin Zheng, PhD[13], Zongyuan Ge, PhD[14], Mingguang He, MD[1,2,15*]

**Affiliations:**

1. School of Optometry, The Hong Kong Polytechnic University, Hong Kong.
2. Research Centre for SHARP Vision (RCSV), The Hong Kong Polytechnic University, Hong Kong.
3. ELLIS Institute Finland, Espoo 02150, Finland.
4. Aalto University, Espoo 02150, Finland.
5. School of Computing, Clemson University, Clemson, SC, USA.
6. Shanghai Eye Diseases Prevention &Treatment Center/ Shanghai Eye Hospital, School of Medicine, Tongji University, Shanghai, China.
7. Department of Ophthalmology, The Second Affiliated Hospital, School of Medicine, Zhejiang University, Hangzhou, China.
8. Centre for Innovation and Precision Eye Health & Department of Ophthalmology, Yong Loo Lin School of Medicine, National University of Singapore, Singapore.
9. Department of Ophthalmology, The University of Hong Kong, Hong Kong.
10. Beijing Tongren Eye Center, Beijing key Laboratory of Intraocular Tumor Diagnosis and Treatment, Beijing Ophthalmology & Visual Sciences Key Laboratory, Medical Artificial Intelligence Research and Verification Key Laboratory of the Ministry of Industry and Information

Technology, Beijing Key Laboratory of Intelligent Diagnosis, Treatment and Prevention of Blinding Eye Diseases, Beijing Tongren Hospital, Capital Medical University, Beijing, China.

11. Department of Ophthalmology and Visual Sciences, The Chinese University of Hong Kong, Hong Kong.
12. State Key Laboratory of Ophthalmology, Zhongshan Ophthalmic Center, Sun Yat-sen University, Guangdong Provincial Key Laboratory of Ophthalmology and Visual Science, Guangdong Provincial Clinical Research Center for Ocular Diseases, Guangzhou, China.
13. Department of Eye and Vision Science, University of Liverpool, Liverpool, UK.
14. AIM for Health Lab, Faculty of Information Technology, Monash University, Melbourne, Victoria, Australia.
15. Centre for Eye and Vision Research (CEVR), 17W Hong Kong Science Park, Hong Kong.

[#] Danli Shi, Xiaolan Chen, Bingjie Yan contributed equally to this article.

**Correspondence:**

*Mingguang He, Department of Experimental Ophthalmology, The Hong Kong Polytechnic University, Hong Kong, China. Email: mingguang.he@polyu.edu.hk

**Running head:** EyeAgent: AI agent for ophthalmology

**Abstract**

**BACKGROUND** Current medical artificial intelligence (AI) systems are often task-specific, poorly integrated, and difficult to interpret, limiting support for multimodal clinical workflows. We developed and evaluated EyeAgent for ophthalmic decision support.

**METHODS** EyeAgent comprises three layers including 54 task-specific EyeTools, 32 clinical skills, and a modular large language model reasoning core with interchangeable backbones. The tool library spans 23 imaging modalities and a 260-condition taxonomy, defining functional scope. A multi-level evaluation distinguished this scope from empirical system-level validation. Component-level evaluation assessed newly developed models on held-out splits of datasets totaling 871,862 samples. System-level evaluation included illustrative cases of workflow breadth, a 200-case expert-rating study of output quality, and an exploratory multicenter reader study comparing unassisted and EyeAgent-assisted performance among 27 ophthalmologists from 11 centers. Quantitative system-level validation focused on specific image-based clinical workflows. Skill-level and backbone ablations further assessed orchestration and robustness.

**RESULTS** In illustrative cases, EyeAgent showed task-specific advantages over GPT-5.5 in lesion quantitation, cross-modality synthesis, visual explanation, evidence retrieval, and uncertainty-aware reporting. In expert rating study, EyeAgent achieved balanced skill selection performance (recall 0.82, precision 0.85, F1 0.84), and 82.7% to 92.0% of outputs were rated acceptable or better across six clinical dimensions. In the exploratory multicenter reader study, diagnostic accuracy (correct or partially correct responses) increased from 72.8% unassisted to 91.9% assisted (difference, 19.1 percentage points [pp]; 95% confidence interval [CI], 13.6-24.7; Holm-adjusted $P<0.001$). Report completeness increased from 55.8% to 75.6% (difference, 19.8 pp; 95% CI, 14.8-24.7; Holm-adjusted $P<0.001$). Skill-level ablation showed that full EyeAgent improved accuracy over the vision-language model-only baseline (76.0% vs. 55.9%), with stable performance across reasoning backbones.

**CONCLUSIONS** EyeAgent provides an agentic AI framework for multimodal ophthalmic analysis and clinical decision support. Across the evaluated settings, it supported interpretable, evidence-guided reasoning and showed potential for human–AI collaboration. Broader real-world effectiveness and generalizability require prospective and cross-population validation.

## Introduction

Artificial intelligence (AI), particularly large language models (LLMs) and vision-language models (VLMs), has advanced medical research.[1-3] In ophthalmology, AI has been applied to screening for vision-threatening diseases,[4] segmenting ocular structure,[5] report generation and interactive question answering.[6] However, most systems address isolated tasks, with limited transparency and adaptability to multimodal workflows requiring evidence integration and stepwise reasoning. Ophthalmology assessment integrates complementary information from imaging modalities including color fundus photography (CFP), optical coherence tomography (OCT), and fundus fluorescein angiography (FFA).[7,8] Clinical reasoning also involves hypothesis testing, cross-modality comparison, longitudinal interpretation, and cross-specialty input to resolve uncertain or conflicting findings, requiring adaptable AI workflows that integrate multimodal inputs and domain-specific tools.

Recent multimodal foundation models show promise but face limitations in interpretability and reliability. End-to-end architectures often produce outputs that are difficult to verify.[9] LLMs are also prone to hallucination, which raises safety concerns in clinical settings.[9,10] Large-scale training also requires annotated datasets constrained by privacy, institutional, and proprietary barriers.[11] Furthermore, many approaches rely on retraining monolithic models, limiting the integration of existing validated tools and external clinical knowledge. Agentic AI offers a promising road,[12,13] by interpreting user requests, planning tasks, retrieving evidence, and orchestrating tools across complex workflows.[14-17] This is particularly relevant to ophthalmology, where validated imaging tools can be organized into clinically meaningful workflows to improve interpretability, evidence integration, and extensibility.

Here we introduce EyeAgent, a three-layer agentic AI system for multimodal ophthalmic decision support. It integrates 54 domain-specific EyeTools (**Fig. 1c**) covering classification, segmentation, generation, biomarker extraction, and retrieval-augmented generation (RAG). A clinical skill library organizes these tools into disease- and scenario-specific workflows covering major diseases such as diabetic retinopathy (DR) and glaucoma risk, as well as broader scenarios such as screening, conflict resolution, and evidence guarding (**Fig. 1b**). An LLM reasoning core coordinates skill selection, tool execution, and evidence integration according to input context (**Fig.1a**), enabling scenario workflows for referral triage, disease-specific diagnosis, quantitative analysis, verification, and evidence-guided recommendation (**Fig. 1d**).

We evaluated EyeAgent through a multi-level framework distinguishing its functional scope from empirical system-level validation using internal and external datasets (**Fig. 1e**). At the component level, new tools were assessed on internal test sets. At the system level, illustrative cases characterized workflow breadth and boundary conditions. Quantitative validation then focused on specific workflows amenable to standardized evaluation, including an external end-to-end expert rating study of 200 cases and an exploratory multicenter human–AI reader study involving 27 ophthalmologists from 11 centers. Skill-level and LLM-backbone ablations assessed orchestration contributions and robustness across reasoning models.

## Methods

### Ethics statement

This study was approved by the Institutional Review Board of The Hong Kong Polytechnic University and conducted in accordance with the Declaration of Helsinki. Only de-identified retrospective data were used, and informed consent was waived.

### EyeAgent system architecture

EyeAgent was designed as a modular three-layer system for multimodal ophthalmic decision support, comprising the EyeTools, clinical skills, and LLM reasoning core (**Fig. 1**). The EyeTools layer integrates 54 task-specific tools, including previously validated and newly developed models, covering classification and detection[18-20], segmentation and quantification[21-25], generation and reporting[6,26-35], and RAG. Tool details are listed in Supplementary Table S1, and datasets in Supplementary Table S2. Each tool is exposed through a standardized Model Context Protocol (MCP)[36] interface, enabling selective invocation and structured outputs. The skill layer contains 32 skills, including 27 disease-specific and 5 cross-cutting skills, that organize tools into clinically meaningful workflows with defined execution sequences, interpretation rules, and evidence safeguards. The LLM reasoning core served as the controller of EyeAgent through a five-stage workflow: (i) query and modality interpretation; (ii) routing preflight and skill selection; (iii) skill-scoped tool execution; (iv) evidence integration and conflict resolution; and (v) structured response generation. The reasoning backbone is interchangeable, with Qwen3.6-27B used as the primary locally deployed model. Further architectural, tool-level, and implementation details are provided in the Supplementary Methods.

## Evaluation of EyeAgent

EyeAgent was retrospectively evaluated using a multi-level framework assessing component readiness and system-level performance in selected clinical workflows (**Fig. 1e**). Newly developed tools were assessed on internal test sets. At the system level, routine and adversarial cases illustrated workflow breadth and boundary conditions, followed by quantitative evaluation through a 200-case expert-rating study and an exploratory multicenter human–AI reader study. The diversity and representativeness of the cases in both studies are summarized in Supplementary Table S3. Skill-level and LLM-backbone ablations further assessed orchestration and robustness.

### *Component-level assessment of newly developed models*

Tool-level validation assessed the 21 newly developed models on predefined test sets (Supplementary Table S2), with evaluation metrics detailed in the Supplementary Methods.

### *System-level qualitative case analysis*

To characterize EyeAgent's workflow breadth and boundary conditions, we conducted a qualitative evaluation using routine and adversarial cases sampled from the independent OphthalWeChat[37] dataset, stratified by disease and imaging modality. Eight routine cases represented distinct clinical workflows, while adversarial cases tested robustness to imaging artifacts and misleading questions. We compared

EyeAgent with GPT-5.5 (without external tools) and an unaided board-certified ophthalmologist (more than 5 years' experience), focusing on task-specific capabilities, hallucinations, and evidence grounding. Detailed case selection and evaluation procedures are provided in the Supplementary Methods.

***System-level end-to-end evaluation***

To assess EyeAgent's end-to-end clinical performance, we evaluated 200 cases spanning 12 ophthalmic conditions (100 CFP and 100 OCT), sampled from OphthalWeChat[37] using disease-stratified random sampling (Supplementary Fig. S1a). EyeAgent autonomously performed routing, skill selection, tool invocation, and diagnostic reporting. Skill selection was evaluated against expert-defined workflows using recall, precision, and F1 score. Two board-certified ophthalmologists independently rated output quality across 6 dimensions: accuracy, completeness, safety, reasoning, interpretability, and evidence grounding, using a 3-point scale (poor, acceptable, good; Supplementary Table S4). Detailed scoring procedures are provided in the Supplementary Methods.

***System-level multicenter human-AI collaboration study***

We conducted an exploratory multicenter reader study involving ophthalmologists from centers in mainland China and Hong Kong using the Qualtrics platform. Junior readers were ophthalmology trainees, whereas senior readers had at least five years of clinical experience. A 30-case pool was constructed from OphthalWeChat[37] using workflow-stratified random sampling and divided into three workflow-matched sets of 10 cases (Supplementary Fig. S1b). Each reader assessed one set in a sequential two-phase, second-reader design: first independently without AI assistance, then after reviewing EyeAgent outputs, with opportunities to revise their diagnoses, reports, and confidence ratings (**Fig.4a; Supplementary Fig. S2**). Readers were blinded to reference diagnoses and AI-output correctness.

The co-primary outcomes were diagnostic accuracy (correct or partially correct responses) and report completeness. Reference diagnoses and report-scoring checklists were established by two independent senior ophthalmologists. Two expert graders independently assessed responses against these standards.

Diagnostic accuracy was determined from their averaged ratings, and report completeness from the mean proportion of applicable checklist items completed. Secondary outcomes included response time, diagnostic confidence, AI adoption, and user feedback. Detailed sampling, grading, and study procedures are provided in the Supplementary Methods.

### *Ablation and benchmarking*

We conducted skill-level and LLM-backbone ablations on the same 200 expert-rating cases. Skill-level ablation compared 5 configurations, from a VLM-only baseline to the full EyeAgent system, plus retrieval and evidence guarding removal, with 5 runs per configuration to assess diagnostic accuracy. For backbone ablation, we compared Qwen3.6-27B, Gemma-4-26B-it, Qwen3.5-9B, GPT-5.5, and DeepSeek V4 Pro using the same EyeAgent workflow (Supplementary Methods).

## Statistical analysis

Descriptive statistics summarized performance metrics. Normalized expert-rating scores were compared between CFP and OCT using the Mann–Whitney U test, with inter-rater reliability assessed by weighted Cohen's kappa or intraclass correlation coefficients. No formal a priori sample-size or power calculation was performed. The exploratory reader study used mixed-effects models accounting for reader and case clustering to analyze diagnostic accuracy, report completeness, response time, and confidence. Holm adjustment was applied to prespecified outcome families. All tests were two-sided ($P<0.05$). Detailed model specifications and sensitivity analyses are provided in the Supplementary Methods.

## Results

### EyeAgent system overview and clinical workflow simulation

EyeAgent integrated 54 ophthalmic tools into a unified agentic framework, including 11 classification and detection tools, 24 segmentation tools, and specialized regression and generation modules. The 21 newly developed models were evaluated on internal test sets, with performance summarized in Supplementary Tables S5–S6. Together, these tools supported 5 clinical functions: referral triage, disease-specific diagnosis, quantitative analysis, verification of uncertain findings, and evidence-guided recommendations. Illustrative cases characterized workflow breadth and boundary conditions (**Fig. 2**), followed by quantitative evaluations of clinical output quality, human–AI collaboration, and system robustness (**Figs. 3–5**).

### Illustrative workflow demonstrations of breadth and boundary conditions

**Figure 2** qualitatively characterized EyeAgent across heterogeneous ophthalmic workflows. In hierarchical decision-making, GPT-5.5 identified the correct disease category but hallucinated modality and laterality, whereas EyeAgent used tool-based metadata extraction and provided confidence scores and lesion visualizations. In quantitative analysis, EyeAgent generated lesion segmentations and measurements from CFP and OCT images. Other examples demonstrated cross-modality synthesis, angiographic reporting, anatomically aligned 3D visualization, retinal vascular biomarker quantification, and evidence-supported recommendations. In adversarial cases (Supplementary Figure S3), EyeAgent used artifact segmentation and multi-tool verification to address misleading image features and user-suggested diagnoses. These comparisons illustrated workflow capabilities and limitations; quantitative performance was assessed separately.

### End-to-end expert rating study

Across 200 cases, EyeAgent achieved a skill-selection recall of 0.82 (CFP 0.83, OCT 0.82), precision of 0.85 (CFP 0.89, OCT 0.81), and F1 score of 0.84 (CFP 0.86, OCT 0.82) (**Fig. 3b**), demonstrating consistent orchestration performance across the evaluated workflows.

In expert evaluations (**Fig. 3c**), the proportion of responses rated as "⩾Acceptable" was 86.5% for accuracy, 92.0% for completeness, and 84.5% for safety. The corresponding proportions were 85.5% for reasoning and 85.0% for interpretability, reflecting coherent clinical reasoning and clear explanations. In cases where evidence citation was applicable (n = 185), citation quality was rated as "⩾Acceptable" in 153 cases (82.7%), suggesting generally reliable evidence integration. Inter-rater reliability was substantial to excellent across the 6 dimensions (weighted Cohen's $\kappa$ = 0.715-0.844). Normalized scores were similar for CFP and OCT (0.795±0.209 vs. 0.830±0.162; P=0.674), and higher for normal retina, glaucoma risk, and DR. A post hoc failure analysis is provided in the Supplementary Results.

**Multicenter human-AI collaboration reader study**

This study included 27 ophthalmologists from 11 centers (18 junior and 9 senior readers; Supplementary Fig. S4). Each reader assessed 10 cases from a workflow-stratified 30-case pool (Fig. 4a). All readers completed both phases for their assigned cases, yielding 270 complete paired reader–case assessments.

EyeAgent assistance was associated with higher diagnostic accuracy and report completeness. Diagnostic accuracy increased from 72.8% unassisted to 91.9% assisted (difference, 19.1 percentage points [pp]; 95% confidence interval [CI], 13.6-24.7; Holm-adjusted P<0.001), while mean report completeness increased from 55.8% to 75.6% (difference, 19.8 pp; 95% CI, 14.8-24.7; Holm-adjusted P<0.001) (Fig.4b). Diagnostic accuracy numerically increased from 68.2% to 91.9% among junior readers and from 82.1% to 91.5% among senior readers, without a significant assistance-by-experience interaction (Holm-adjusted P=0.087). Report completeness increased from 51.3% to 74.9% among junior readers and from 64.7% to 76.9% among senior readers, with a greater assistance effect among junior readers (Holm-adjusted interaction P=0.013) (Fig. 4c; Supplementary Table S7). Sensitivity analyses showed consistent findings (Supplementary Table S8). EyeAgent alone achieved 96.7% diagnostic accuracy (29/30; 95% CI, 83.3–99.4%) and a mean report completeness score of 0.81±0.16.

The assisted revision phase was shorter than the initial unassisted assessment by 147.5 seconds (s) overall (100.7s among senior readers and 167.0s among junior readers; Fig. 4d; Supplementary Table S9). Decision confidence increased by 0.51 points overall (0.19 points among senior readers and 0.67 points among junior readers; Fig. 4e; Supplementary Table S9). Readers selectively adopted EyeAgent outputs (Fig. 4f; Supplementary Table S10), with 75% adoption most frequently observed (30.7%, 83/270) and similar adoption distributions across experience groups. Adopted components included management recommendations (28.1%), diagnostic evidence (24.0%), diagnosis (23.3%), and image details (20.5%). Post-task feedback showed high acceptance (Fig. 4g), with 88.9% of participants (24/27 for each item) reporting satisfaction, clear reasoning, and willingness to use similar AI support in clinical practice or training.

Collectively, EyeAgent assistance was associated with improved diagnostic accuracy, report completeness, and confidence, with greater completeness gains among junior readers.

**Ablation and robustness analysis**

Across repeated evaluations (**Fig. 5a**), diagnostic accuracy increased from 55.9 ± 1.1% with the VLM-only baseline to 76.0 ± 0.5% with full EyeAgent. Accuracy reached 63.5 ± 1.8% with all tools but no skills, 74.7 ± 0.7% with routing preflight, and 75.6 ± 1.0% after adding disease-specific workflow skills, supporting the contribution of skill-guided orchestration. Removing retrieval or evidence guarding reduced accuracy to 75.4 ± 0.7% and 75.1 ± 0.7%, respectively, suggesting additional contributions from evidence integration and verification.

Backbone ablation using the same full-skill workflow (**Fig. 5b**) showed comparable accuracy across models, with 76.5% for Qwen3.6-27B and Gemma-4-26B-it, 77.0% for GPT-5.5, and 73.0% for DeepSeek V4 Pro and Qwen3.5-9B. The 0.5-pp difference between Qwen3.6-27B and the highest-performing model indicated relatively stable performance across backbones. Qwen3.6-27B was selected as the primary backbone for its competitive accuracy and local deployment capabilities, supporting medical data privacy, infrastructure control, and more predictable operating costs than externally hosted models.

Across 200 cases, mean latency was 102.44 seconds per case (median, 76.26 seconds), corresponding to an estimated serial throughput of 35.1 cases per hour. Mean token usage was 34,924 per case (27,744 prompt and 7,180 completion tokens). Dynamic GPU allocation kept EyeTools memory usage below 40 GB without loading all tools simultaneously.

## Discussion

We present EyeAgent, a skill-based explainable and extensible framework integrating 32 ophthalmic skills and 54 specialized tools into structured clinical workflows. Illustrative cases characterized workflow breadth, while quantitative system-level evaluation showed effective orchestration and favorable expert-rated output quality in the evaluated simulated workflows. In the exploratory multicenter reader study, EyeAgent assistance was associated with higher diagnostic accuracy and report completeness and shorter response times, with increased confidence particularly among junior readers.

Agentic AI enables autonomous planning, reasoning, and tool use in medicine. Prior work such as MMedAgent and task-specific LLM-tool hybrids have shown the benefits of tool integration over standalone VLMs.[38,39] EyeAgent extends tool integration through disease- and scenario-specific skills coordinating routing, tool execution, and evidence synthesis. Such an approach may be particularly useful for tasks requiring multimodal interpretation, quantitative analysis, and stepwise decision support.

Transparent, evidence-based decision-making is essential in clinical medicine.[40] EyeAgent supports transparency by stepwise reasoning and traceable outputs, including lesion maps, biomarker measurements, invoked tools, retrieved evidence, and uncertainty caveats, allowing clinicians to inspect its recommendations. Illustrative cases showed how quantitative measurements and retrieved evidence complemented clinical judgment. Consistently, experts rated reasoning, interpretability, and evidence grounding favorably, while readers reported high satisfaction and willingness to use EyeAgent, suggesting that transparency and inspectability may facilitate clinician acceptance.

Hallucination remains a major limitation of LLMs in medicine. EyeAgent incorporates tool-grounded reasoning, retrieval, and conflict-resolution mechanisms to reduce unsupported outputs. In illustrative comparisons with GPT-5.5, metadata extraction, specialized tools, and retrieved evidence helped address incorrect or unsupported interpretations, while additional analysis or retrieval enabled verification of uncertain or conflicting outputs. These evidence-based mechanisms are consistent with favorable expert ratings for reasoning, safety, interpretability, and evidence grounding.

Modularity is a key advantage of agentic systems. Ablation results showed that skill-guided routing and disease-specific workflows improved performance beyond tool availability alone, with additional contributions from retrieval and evidence guarding. Importantly, EyeAgent's modular design enables flexible integration of new algorithms, updated tools, regulatory-approved devices, and subspecialty modules. Illustrative anterior-segment cases demonstrated this extensibility beyond retinal workflows, although dedicated validation remains necessary. Modularity further extends to the reasoning backbone, with comparable performance across models allowing deployment to balance accuracy, privacy, latency, and cost. This flexibility supports locally hosted models for data privacy, while smaller backbones and selective tool loading may facilitate deployment in resource-limited clinics, subject to further validation.

The retina serves as a "window" to systemic health, with retinal age and vascular features emerging as biomarkers of systemic disease.[41-44] EyeAgent could integrate ophthalmic and systemic insights through cross-specialty multi-agent collaboration to support more comprehensive health assessments.

This work has several limitations. First, EyeAgent was evaluated primarily as an agentic framework. Although its library spans 23 imaging modalities and a 260-condition screening taxonomy, quantitative system-level validation focused mainly on selected CFP- and OCT-based, predominantly retinal workflows. The findings support framework feasibility, orchestration, and clinical output quality in these settings but cannot establish performance across all supported tools, modalities, and diseases. Anterior-segment and longitudinal cases illustrated broader capabilities, requiring dedicated prospective validation.

Second, EyeAgent's extensive tool library introduces performance heterogeneity, with overall reliability dependent on individual components. Errors in image-quality assessment, disease screening, segmentation, or retrieval may propagate into final outputs, requiring module-level optimization, independent external validation, and retrieval-quality monitoring. Furthermore, EyeAgent is not yet production-ready; clinical deployment requires integration with structured clinical data and hospital information systems, privacy safeguards, and regulatory review.

Third, expert-rated safety reflected retrospective clinical acceptability without systematically assessing adverse events, error propagation, or clinician reliance, which require prospective evaluation before clinical deployment. Glaucoma-related outputs were limited to CFP-based risk assessment; more comprehensive evaluation should incorporate intraocular pressure, visual fields, retinal nerve fiber layer thickness, and optic disc findings.

Finally, the exploratory reader study lacked a formal a priori power calculation. The curated OphthalWeChat cases and disease- and workflow-stratified sampling provided structured coverage but may not reflect real-world disease prevalence or clinical case mix. Accordingly, the reported estimates characterize the constructed evaluation cohorts and should not be interpreted as prevalence-dependent estimates for routine clinical practice. In addition, the convenience sample of readers from mainland China and Hong Kong was not matched to cases by subspecialty expertise, further limiting generalizability. The fixed-order design could not distinguish AI assistance effects from order effects, and shorter assisted revision times do not establish reduced total workflow time. Larger prospectively powered studies using consecutive real-world cases and randomized crossover or counterbalanced designs are needed to assess generalizability and estimate the independent effect of AI assistance.

In conclusion, EyeAgent provides a modular agentic AI framework combining LLM reasoning, extensible skills, and multimodal tools into structured ophthalmic workflows. It showed workflow flexibility across tasks, while quantitative evaluation supported clinically organized orchestration and potential clinician assistance within evaluated settings. Further prospective validation across broader modalities, diseases, subspecialties, and clinical populations is needed to establish generalizability and guide responsible clinical translation.

**Reference**

1. Rajpurkar P, Chen E, Banerjee O, Topol EJ. AI in health and medicine. Nature medicine 2022;28(1):31-38.

2. Thirunavukarasu AJ, Ting DSJ, Elangovan K, Gutierrez L, Tan TF, Ting DSW. Large language models in medicine. Nature Medicine 2023;29(8):1930-1940. (In eng). DOI: 10.1038/s41591-023-02448-8.

3. Shi D, Zhang W, Yang J, et al. A multimodal visual–language foundation model for computational ophthalmology. npj Digital Medicine 2025;8(1):381. DOI: 10.1038/s41746-025-01772-2.

4. Gulshan V, Peng L, Coram M, et al. Development and Validation of a Deep Learning Algorithm for Detection of Diabetic Retinopathy in Retinal Fundus Photographs. JAMA 2016;316(22):2402-2410. (In eng). DOI: 10.1001/jama.2016.17216.

5. Ting DSW, Pasquale LR, Peng L, et al. Artificial intelligence and deep learning in ophthalmology. British Journal of Ophthalmology 2019;103(2):167. DOI: 10.1136/bjophthalmol-2018-313173.

6. Chen X, Zhang W, Xu P, et al. FFA-GPT: an automated pipeline for fundus fluorescein angiography interpretation and question-answer. npj Digital Medicine 2024;7(1):111. DOI: 10.1038/s41746-024-01101-z.

7. He M, Li Z, Liu C, Shi D, Tan Z. Deployment of artificial intelligence in real-world practice: opportunity and challenge. Asia-Pacific Journal of Ophthalmology 2020;9(4):299-307.

8. Cheng C-Y, Soh ZD, Majithia S, et al. Big Data in Ophthalmology. Asia-Pacific Journal of Ophthalmology (Philadelphia, Pa) 2020;9(4):291-298. (In eng). DOI: 10.1097/APO.0000000000000304.

9. Huang L, Yu W, Ma W, et al. A survey on hallucination in large language models: Principles, taxonomy, challenges, and open questions. ACM Transactions on Information Systems 2025;43(2):1-55.

10. Kwong JCC, Wang SCY, Nickel GC, Cacciamani GE, Kvedar JC. The long but necessary road to responsible use of large language models in healthcare research. NPJ Digit Med 2024;7(1):177. (In eng). DOI: 10.1038/s41746-024-01180-y.

11. Zhang K, Zhou R, Adhikarla E, et al. A generalist vision–language foundation model for diverse biomedical tasks. Nature Medicine 2024:1-13.

12. Durante Z, Huang Q, Wake N, et al. Agent AI: Surveying the Horizons of Multimodal Interaction. CoRR 2024.

13. Chen X, Chen R, Xu P, et al. From visual question answering to intelligent AI agents in ophthalmology. British Journal of Ophthalmology 2025:bjo-2024-326097. DOI: 10.1136/bjo-2024-326097.

14. Qiu J, Lam K, Li G, et al. LLM-based agentic systems in medicine and healthcare. Nat Mach Intell 2024;6(12):1418-1420.

15. Gao S, Fang A, Huang Y, et al. Empowering biomedical discovery with AI agents. Cell 2024;187(22):6125-6151. DOI: 10.1016/j.cell.2024.09.022.

16. de Almeida BP, Richard G, Dalla-Torre H, et al. A multimodal conversational agent for DNA, RNA and protein tasks. Nature Machine Intelligence 2025:1-14.

17. Cvpr. Drug Discovery Agent: An Automated Vision Detection System For Drug-Cell Interactions.

18. Li Z, Keel S, Liu C, et al. An Automated Grading System for Detection of Vision-Threatening Referable Diabetic Retinopathy on the Basis of Color Fundus Photographs. Diabetes Care 2018;41(12):2509-2516. (In eng). DOI: 10.2337/dc18-0147.

19. Keel S, Li Z, Scheetz J, et al. Development and validation of a deep-learning algorithm for the detection of neovascular age-related macular degeneration from colour fundus photographs. Clinical & Experimental Ophthalmology 2019;47(8):1009-1018. (In eng). DOI: 10.1111/ceo.13575.

20. Li Z, He Y, Keel S, Meng W, Chang RT, He M. Efficacy of a Deep Learning System for Detecting Glaucomatous Optic Neuropathy Based on Color Fundus Photographs. Ophthalmology 2018;125(8):1199-1206. (In eng). DOI: 10.1016/j.ophtha.2018.01.023.

21. He S, Joseph S, Bulloch G, et al. Bridging the Camera Domain Gap With Image-to-Image Translation Improves Glaucoma Diagnosis. Translational Vision Science & Technology 2023;12(12):20-20. DOI: 10.1167/tvst.12.12.20.

22. Shi D, Lin Z, Wang W, et al. A Deep Learning System for Fully Automated Retinal Vessel Measurement in High Throughput Image Analysis. Front Cardiovasc Med 2022;9:823436. DOI: 10.3389/fcvm.2022.823436.

23. Chen R, Zhao Z, Yusufu M, Shang X, Shi D, He M. Choroidal Vessel Segmentation on Indocyanine Green Angiography Images via Human-in-the-Loop Labeling. arXiv preprint arXiv:240601993 2024.

24. Zhao Z, Huang S, Zhang W, et al. Automated Interpretation of Fundus Fluorescein Angiography with Multi-Retinal Lesion Segmentation. medRxiv 2024:2024.12. 20.24319428.

25. Zhang W, Tian Z, Song F, Xu P, Shi D, He M. Enhancing stability in cardiovascular disease risk prediction: A deep learning approach leveraging retinal images. Informatics in Medicine Unlocked 2023;42:101366. DOI: 10.1016/j.imu.2023.101366.

26. Shi D, Zhang W, He S, et al. Translation of Color Fundus Photography into Fluorescein Angiography Using Deep Learning for Enhanced Diabetic Retinopathy Screening. Ophthalmol Sci 2023;3(4):100401. DOI: 10.1016/j.xops.2023.100401.

27. Chen R, Zhang W, Song F, et al. Translating color fundus photography to indocyanine green angiography using deep-learning for age-related macular degeneration screening. NPJ Digit Med 2024;7(1):34. (In eng). DOI: 10.1038/s41746-024-01018-7.

28. Song F, Zhang W, Zheng Y, Shi D, He M. A deep learning model for generating fundus autofluorescence images from color fundus photography. Adv Ophthalmol Pract Res 2023;3(4):192-198. DOI: 10.1016/j.aopr.2023.11.001.

29. Chen R, Zhang W, Liu B, et al. EyeDiff: text-to-image diffusion model improves rare eye disease diagnosis. arXiv preprint arXiv:241110004 2024.

30. Wu X, Wang L, Chen R, et al. Generation of Fundus Fluorescein Angiography Videos for Health Care Data Sharing. JAMA Ophthalmology 2025. DOI: 10.1001/jamaophthalmol.2025.1419.

31. Chen X, Zhang W, Zhao Z, et al. ICGA-GPT: report generation and question answering for indocyanine green angiography images. The British Journal of Ophthalmology 2024;108(10):1450-1456. (In eng). DOI: 10.1136/bjo-2023-324446.

32. Zhao Z, Zhang W, Chen X, et al. Slit Lamp Report Generation and Question Answering: Development and Validation of a Multimodal Transformer Model with Large Language Model Integration. Journal of Medical Internet Research 2024;26:e54047. (In eng). DOI: 10.2196/54047.

33. Zhang W, Huang S, Yang J, et al. Fundus2Video: Cross-Modal Angiography Video Generation from Static Fundus Photography with Clinical Knowledge Guidance. Medical Image Computing and Computer Assisted Intervention – MICCAI. Morocco: Springer Nature Switzerland; 2024:689-699.

34. Shi D, Liu B, Tian Z, et al. Fundus2Globe: Generative AI-Driven 3D Digital Twins for Personalized Myopia Management. arXiv preprint arXiv:250213182 2025.

35. Liu B, Zhang W, Chotcomwongse P, et al. APTOS-2024 challenge report: Generation of synthetic 3D OCT images from fundus photographs. arXiv preprint arXiv:250607542 2025.

36. Anthropic. 2024. Introducing the Model Context Protocol. (https://www.anthropic.com/news/model-context-protocol.).

37. Xu P, Gong X, Chen X, et al. Benchmarking Large Multimodal Models for Ophthalmic Visual Question Answering with OphthalWeChat. Advances in Ophthalmology Practice and Research 2025. DOI: https://doi.org/10.1016/j.aopr.2025.10.006.

38. Li B, Yan T, Pan Y, et al. Mmedagent: Learning to use medical tools with multi-modal agent. arXiv preprint arXiv:240702483 2024.

39. Goodell AJ, Chu SN, Rouholiman D, Chu LF. Large language model agents can use tools to perform clinical calculations. NPJ Digit Med 2025;8(1):163. (In eng). DOI: 10.1038/s41746-025-01475-8.

40. Reddy S. Explainability and artificial intelligence in medicine. Lancet Digit Health 2022;4(4):e214-e215. (In eng). DOI: 10.1016/s2589-7500(22)00029-2.

41. Nusinovici S, Rim TH, Yu M, et al. Retinal photograph-based deep learning predicts biological age, and stratifies morbidity and mortality risk. Age and Ageing 2022;51.

42. Shi D, Zhou Y, He S, et al. Cross-modality Labeling Enables Noninvasive Capillary Quantification as a Sensitive Biomarker for Assessing Cardiovascular Risk. Ophthalmol Sci 2024;4(3):100441. DOI: 10.1016/j.xops.2023.100441.

43. Yusufu M, Friedman DS, Kang M, et al. Retinal vascular fingerprints predict incident stroke: findings from the UK Biobank cohort study. Heart (British Cardiac Society) 2025;111(7):306-313. (In eng). DOI: 10.1136/heartjnl-2024-324705.

44. Li C, Huang Y, Chen J, et al. Retinal oculomics and risk of incident aortic aneurysm and aortic adverse events: a population-based cohort study. Int J Surg 2025 (In eng). DOI: 10.1097/js9.0000000000002236.

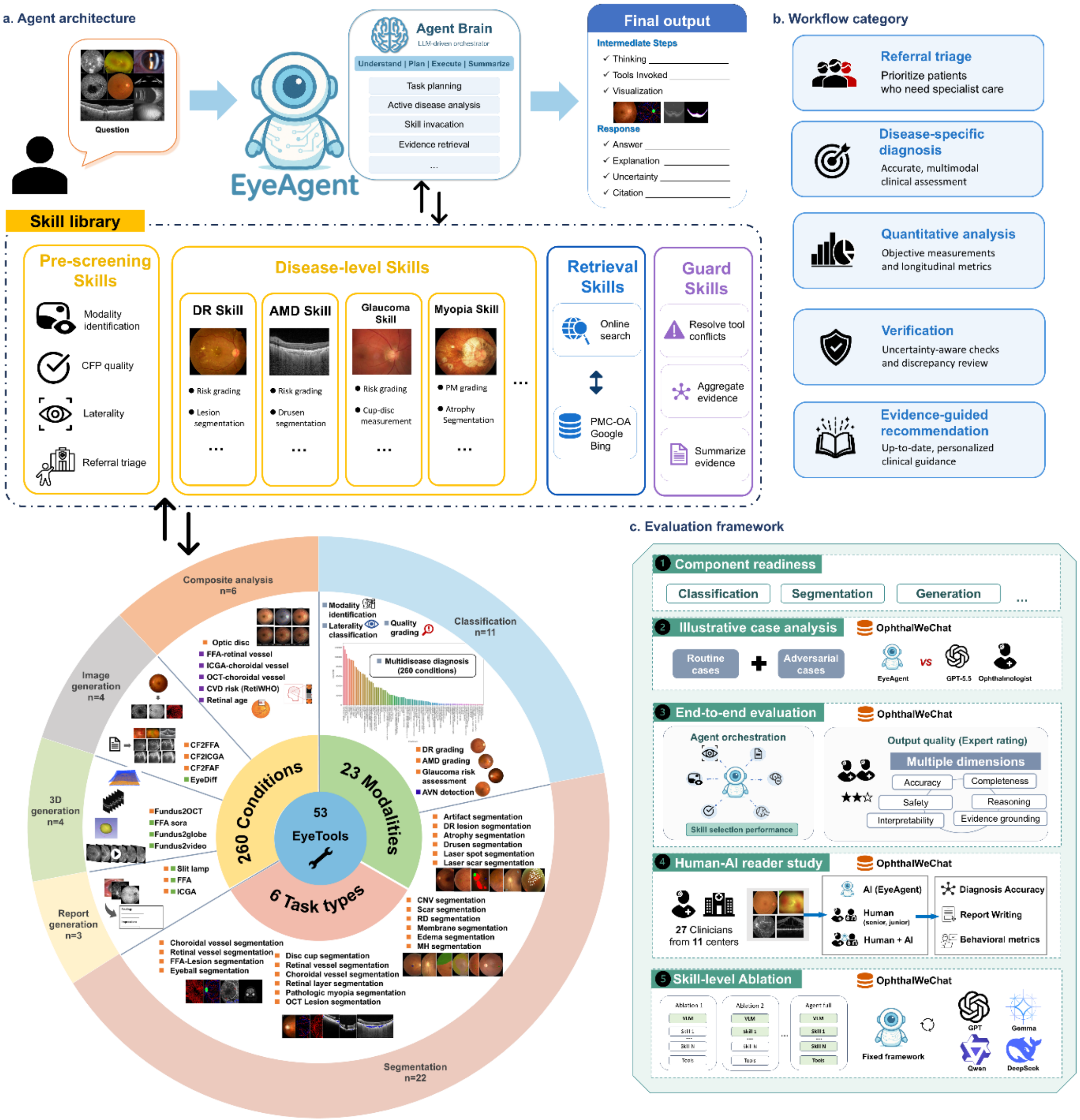

**Figure 1. Overview of EyeAgent system and study design.**

a, Agent architecture. The system receives free-text clinical questions and ophthalmic images, interprets the task, plans the workflow, invokes relevant skills and tools, and generates structured outputs with intermediate steps, tool-use records, visualizations, explanations, uncertainty statements, and citations. The EyeSkill library is organized into pre-screening skills, disease-level skills, retrieval skills, and guard skills. Pre-screening skills support modality identification, image-quality assessment, laterality detection, and referral triage. Disease-level skills organize disease-specific workflows such as diabetic retinopathy and age-related macular degeneration. Retrieval and guard skills support evidence retrieval, conflict resolution, evidence aggregation, and evidence summarization. The EyeTools library integrates 54 ophthalmic tools covering classification, segmentation, regression and analysis, report generation, image generation, 3D generation, and retrieval-augmented generation. Their interfaces span 23 imaging modalities and a 260-condition taxonomy, defining the technical scope of the current system.

b, Workflow categories supported by EyeAgent, including category. including referral triage, disease-specific diagnosis, quantitative analysis, verification, and evidence-guided recommendation.

c, Evaluation framework. The evaluation framework comprised three levels: technical scope, component-level readiness, and system-level empirical validation. Component-level assessment evaluated newly developed tools; system-level evaluation included illustrative case analysis, an end-to-end expert-rating study, an exploratory multicenter human–AI reader study, and skill-level and backbone ablation analyses. Quantitative system-level validation was conducted within prespecified tested workflows.

Abbreviations: LLM, large language model; CFP, color fundus photography; FFA, fundus fluorescein angiography; OCT, optical coherence tomography; ICGA, indocyanine green angiography; SLO, scanning laser ophthalmoscopy; DR, diabetic retinopathy; AMD, age-related macular degeneration; PM, pathologic myopia; PMC-OA, PubMed Central Open Access.

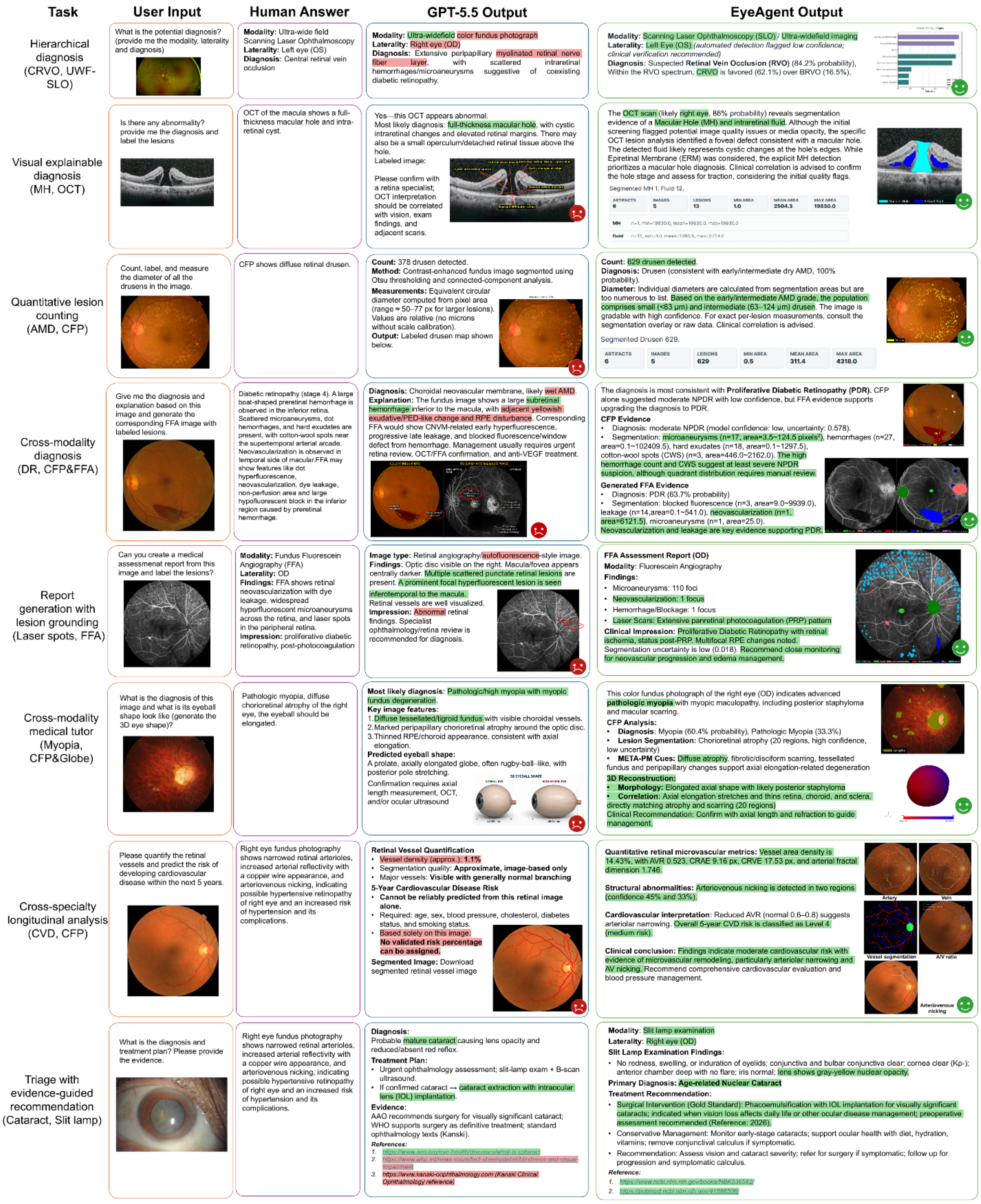

**Figure 2. Illustrative workflow demonstrations of EyeAgent across routine clinical workflows.**

Representative routine cases comparing the outputs of an unaided ophthalmologist, GPT-5.5, and EyeAgent across multiple ophthalmic tasks. The examples include hierarchical diagnosis, visual explainable diagnosis, quantitative lesion counting, cross-modality diagnosis, report generation with lesion grounding, cross-modality medical tutoring, cross-specialty longitudinal analysis, and triage with evidence-guided recommendation. Correct or clinically supported statements are highlighted in green, whereas incorrect or unsupported statements are highlighted in red. Desired and undesired visual outputs are denoted by smile and frown icons, respectively.

Abbreviations: CRVO, central retinal vein occlusion; UWF-SLO, ultra-widefield scanning laser ophthalmoscopy; MH, macular hole; OCT, optical coherence tomography; AMD, age-related macular degeneration; CFP, color fundus photography; FFA, fundus fluorescein angiography; DR, diabetic retinopathy; CVD, cardiovascular disease; NPDR, non-proliferative diabetic retinopathy; PDR, proliferative diabetic retinopathy; IOL, intraocular lens.

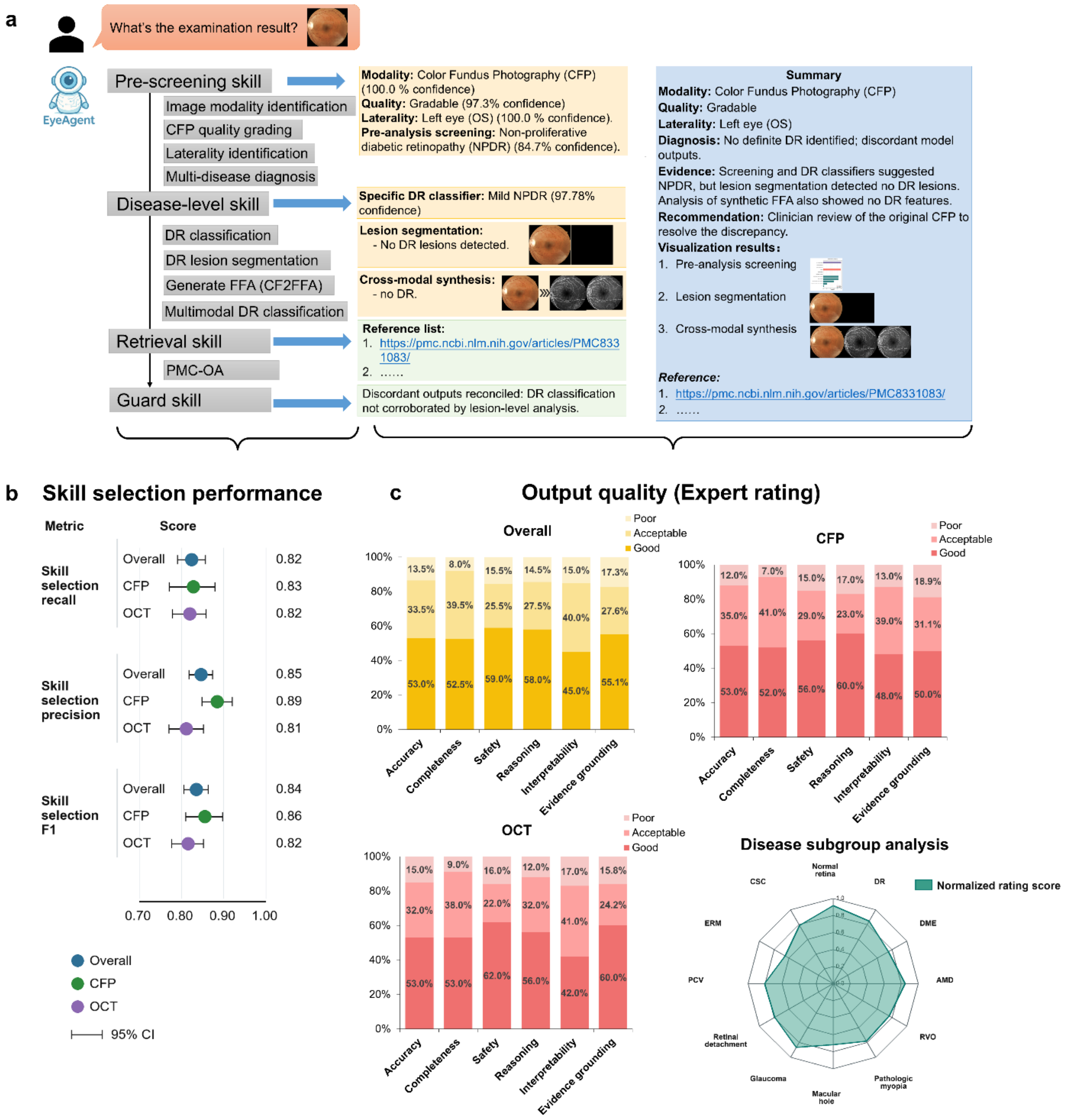

**Figure 3. End-to-end expert evaluation of EyeAgent orchestration and output quality.**

a, Representative example of EyeAgent's end-to-end workflow. After receiving an input image and clinical question, EyeAgent performs pre-screening, selects disease-level skills, invokes retrieval and guard skills when needed, and generates a final response with diagnosis, supporting evidence, recommendations, visualization results, and references.

b, Skill selection performance across 200 cases from OphthalWeChat, summarized by recall, precision, and F1 score for the overall cohort and CFP and OCT subgroups. Error bars indicate 95% confidence intervals.

c, Expert ratings of EyeAgent output quality across 6 clinical dimensions: accuracy, completeness, safety, reasoning, interpretability, and evidence grounding. Stacked bars show the proportions of outputs rated as Poor, Acceptable, or Good overall and in CFP and OCT subgroups. Disease subgroup values represent the mean of case-level normalized overall scores, calculated by dividing the retained total rating by the maximum possible score across applicable dimensions (18 for six dimensions or 15 for five dimensions).

Abbreviations: CFP, color fundus photography; OCT, optical coherence tomography; DR, diabetic retinopathy; DME, diabetic macular edema; AMD, age-related macular degeneration; RVO, retinal vein occlusion; CSC, central serous chorioretinopathy; ERM, epiretinal membrane; PCV, polypoidal choroidal vasculopathy.

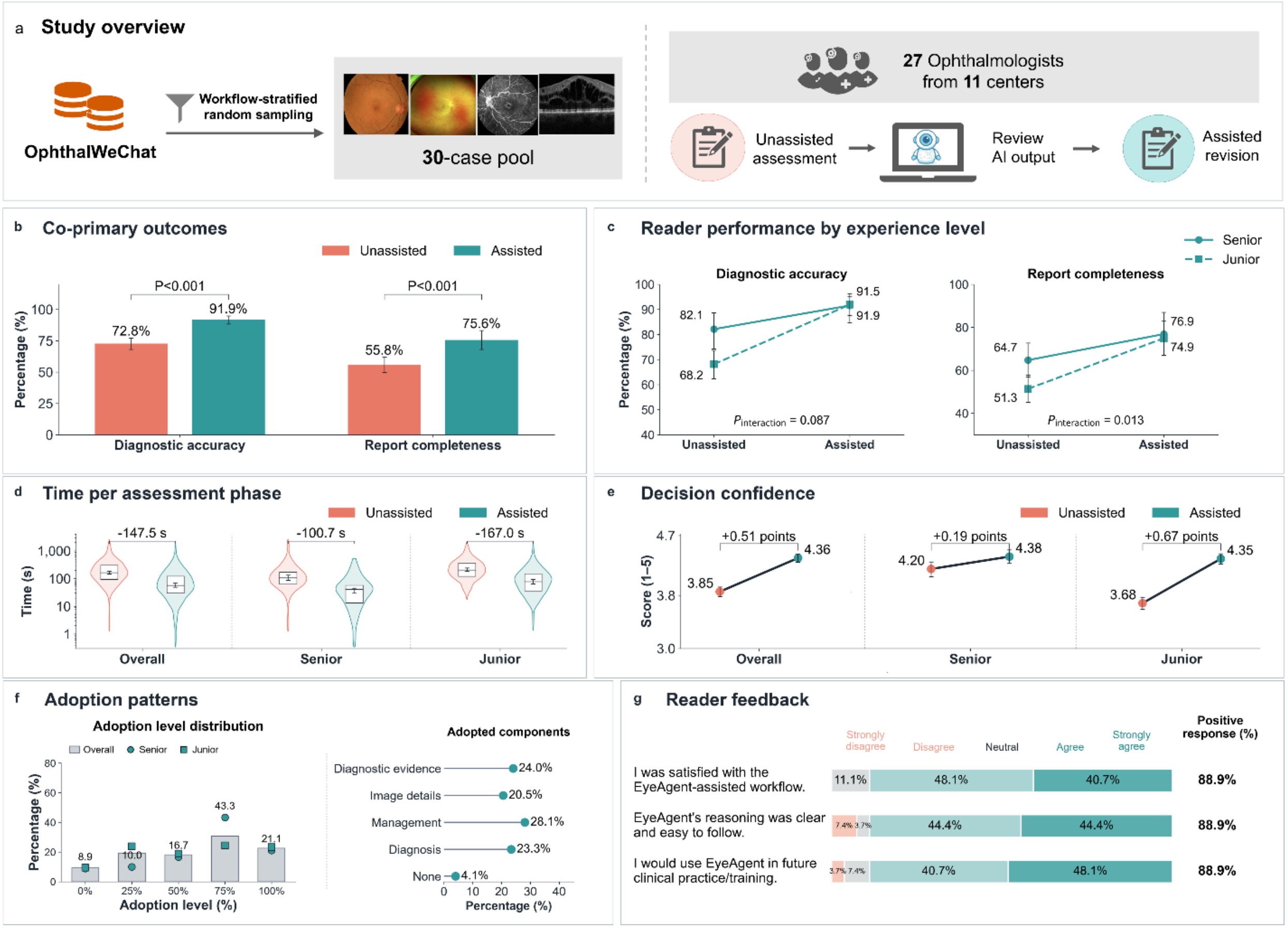

**Figure 4. Human–AI collaboration reader study evaluating EyeAgent.**

a, Study overview. The multicenter study involved 27 ophthalmologists from 11 centers, each reviewing 10 cases sampled from a 30-case pool, first without AI assistance and then after reviewing EyeAgent output, yielding 270 paired reader–case assessments, comprising 540 phase-specific observations.

b, Co-primary outcomes. Diagnostic accuracy includes correct or partially correct responses and is summarized using a logistic model with crossed reader and case random intercepts. Report completeness is the mean of the two graders' proportions for each report, summarized using a Gaussian mixed model with reader, case, and reader–case random intercepts and reader and case random assistance slopes. Bars show model-based means and error bars indicate 95% intervals. P values were Holm-adjusted across the two co-primary outcomes.

c, Diagnostic accuracy and report completeness by reader experience. Points show model-based means with 95% intervals from models including assistance-by-experience interactions. Interaction P values were Holm-adjusted across the two co-primary outcomes; the diagnostic interaction is on the log-odds scale and the completeness interaction is on the mean-difference scale.

d, Response-time distributions for unassisted and assisted assessments, shown on a logarithmic scale. Violin plots show observed distributions, and box plots indicate the median and interquartile range. Values above each pair indicate model-adjusted differences estimated from crossed random-effects models with reader and case as random effects.

e, Decision confidence before and after EyeAgent review. Points show model-adjusted estimates from crossed random-effects models with 95% CI; values above each pair indicate the adjusted change in confidence score.

f, Descriptive adoption patterns during the assisted phase, including reported adoption levels and components of EyeAgent output incorporated by readers. For adopted components, percentages represent the share of all selected components; multiple components could be selected per reader–case observation, and None indicates that no component was adopted.

g, Descriptive reader feedback after study completion using a 5-point Likert scale. Positive responses combine agree and strongly agree.

CI, confidence interval.

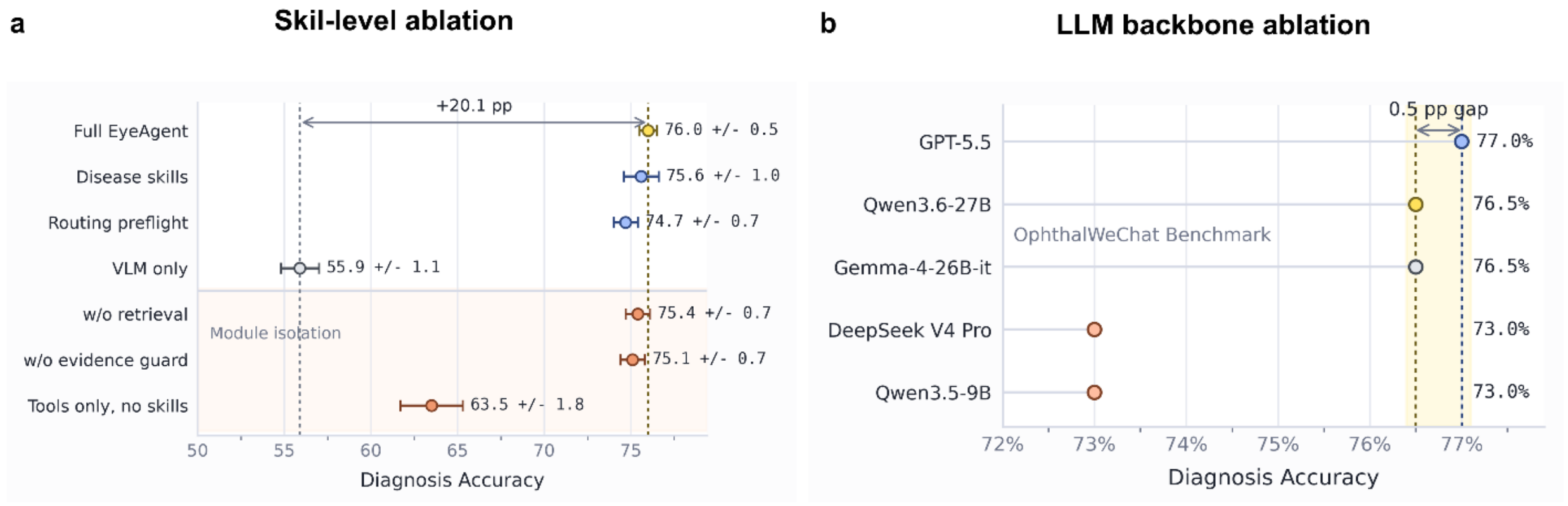

**Figure 5. Ablation and backbone robustness analyses.**

a, Skill-level ablation on the 200-case OphthalWeChat benchmark. Diagnostic accuracy was compared across a VLM-only baseline, routing-preflight configuration, disease-skill configuration, full EyeAgent, and module-isolation settings with online retrieval or evidence-guard constraints removed. The all-tools/no-skills setting tested whether direct tool exposure without skill-guided workflow organization was sufficient. Values are shown as mean ± SD across repeated runs.

b, LLM backbone ablation using the same full EyeAgent workflow and fixed tool registry, skill prompts, and evaluation protocol. Diagnostic accuracy was compared across GPT-5.5, Qwen3.6-27B, Gemma-4-26B-it, DeepSeek V4 Pro, and Qwen3.5-9B. The dotted line indicates the 200-case benchmark reference for the default full-skill workflow.

Abbreviations: VLM, vision-language model; LLM, large language model; SD, standard deviation; pp, percentage points.